\documentclass[osajnl,twocolumn,showpacs,superscriptaddress,10pt,nofootinbib]{revtex4-1} 
\usepackage{amsmath,amssymb,graphicx,verbatim}
\usepackage{bbold,color} 
\usepackage[normalem]{ulem}

\usepackage[margin=1in]{geometry}
\newcommand{\comments}[1]{}
\def\({\left(}
\def\){\right)}

\def\d#1{#1^\dagger}

\newcommand\fig[1]{Fig.~\ref{fig:#1}}

\newcommand\Fig[1]{Figure~\ref{fig:#1}}
\newcommand\tb[1]{Table~\ref{tab:#1}}

\newcommand\bra[1]{\left\langle\,#1\,\right|} 
\newcommand\ket[1]{\left|\,#1\,\right\rangle}

\graphicspath{{Graphics/}}

\begin{document}

\title{Direct measurement of the Wigner function by photon-number-resolving detection}
\author{Niranjan Sridhar}\email{Corresponding author: ns4mf@virginia.edu}
\author{Reihaneh Shahrokhshahi}
\affiliation{University of Virginia, Dept.\ of Physics, 382 McCormick Rd., Charlottesville, VA 22903, USA}

\author{Aaron J. Miller}
\affiliation{Albion College, 611 E. Porter St., Albion, MI 49224, USA}

\author{Brice Calkins}
\author{Thomas Gerrits}
\author{Adriana Lita}
\author{Sae Woo Nam}
\affiliation{National Institute of Standards and Technology, 325 Broadway, Boulder, CO 80303, USA}

\author{Olivier Pfister}
\affiliation{University of Virginia, Dept.\ of Physics, 382 McCormick Rd., Charlottesville, VA 22903, USA}

\begin{abstract}
Photon-number-revolving (PNR) detection allows the direct measurement of the Wigner quasiprobability distribution of an optical mode without the need for numerically processing an inverse Radon transform [K. Banaszek and K. W\'odkiewicz, Phys.\ Rev.\ Lett.\ {\bf 76}, 4344 (1996)]. In this work, we reproduced the seminal experiment of Banaszek et al.\ [Phys.\ Rev.\ A {\bf 60}, 674 (1999)] of quantum tomography of a pure coherent state, and of a statistical mixture thereof, and extended it to the more general case of photon fluxes with much more than one photon per detection time. This was made possible by the use of a superconducting transition-edge sensor to perform PNR detection from 0 to 5 photons at 1064 nm, at $\sim70$\% system efficiency and with no dead time. We detail signal acquisition and detection efficiency and discuss prospects for applying such quantum tomography to non-Gaussian states.
\end{abstract}


\maketitle 

\section{Introduction}

The complete characterization of the quantum state, a.k.a.\ quantum state tomography or quantum tomography, of a physical system plays a key role in physics, in particular in quantum information. Early on, optical systems naturally lent themselves to quantum tomography, likely due to the classical prevalence of their wave nature. The particular case of the continuous variables constituted by the canonically conjugate amplitudes of the quantized electromagnetic field is well known. In this case, access is gained to the amplitude and phase quadrature field operators, $Q=(a+a^{\dag})/\sqrt2$ and $P=i(a^{\dag}-a)/\sqrt2$, by way of homodyne detection of this quantum field with a local oscillator --- in practice a well stabilized laser field. The measurement histograms of a sufficient number of rotated field quadratures represent marginal probability distributions of the Wigner quasiprobability distribution~\cite{Wigner1932}, which can then be reconstructed using numerical inverse Radon transform postprocessing~\cite{Smithey1993,Leonhardt1997,Lvovsky2009}.

An alternate and more direct method to measure the Wigner function, free of the encumbrance of the reconstruction process~\cite{Blume-Kohout2010}, was proposed by Banaszek and W\'odkiewicz~\cite{Banaszek1996}, and also by Wallentowitz and Vogel~\cite{Wallentowitz1996}. This method relies on ideal photon counting, a.k.a.\ photon-number-resolving (PNR) detection, defined as an ideal measurement in the Fock state basis. 

{This requires the experimental capability of photon-number resolution, which also implies, in principle, unity quantum detection efficiency since Fock states form an orthogonal basis. It is important to note, however, that the effect of losses (and thus of nonideal detection efficiency) can be, in principle, deconvoluted exactly in the general case if the overall efficiency is larger than 50\%~\cite{Kiss1995}, and at the cost of taking additional data. In the particular case of states whose nature is invariant under losses (e.g., nonsqueezed Gaussian states such as coherent states), Bondani et al.\ showed that linear photodetectors with intrinsic gain can be used to perform analysis of classical states~\cite{Bondani2009,Bondani2010}.}

{The arduous experimental requirement of achieving PNR was circumvented by use of detectors with single-photon sensitivity, such as photomultipliers or avalanche photodiodes, both in the Geiger mode. Since such detectors do not, in most cases, resolve the photon number, one also has to work at photon fluxes low enough that there would be a negligible probability of more than one photon in the detection's temporal window. Under such conditions, Banaszek et al.\ demonstrated coherent-state quantum tomography~\cite{Banaszek1999a}, with an effective restriction to the $\{\ket0,\ket1\}$ subset of the Fock basis. More recently, the quantum properties of pulsed light fields were also investigated point by point in phase space~\cite{Laiho2010}. When the input has higher photon flux, one can also reach PNR by splitting a multiphoton input~\cite{Song1990} onto efficient single-photon detectors whose count is integrated~\cite{Achilles2003,Fitch2003}.
}
 
{Our ultimate goal is to achieve unconditional direct-detection state tomography using highly efficient PNR detectors. As a preliminary demonstration,} we reproduced Banaszek et al.'s seminal experiment 
{on coherent states with} a superconducting transition-edge sensor (TES), of system detection efficiency above 90\%~\cite{Lita2008}. These detectors can distinguish between 0- to 5-photon Fock states at 1064 nm with high system detection efficiency, no dead time, and near zero dark count. {Recently, the TES was quantum characterized to be a linear detector~\cite{Brida2012}.} Therefore the TES can measure the state without any fair sampling assumption. We believe this to be a step towards more direct state reconstruction of nonclassical states (i.e., with minimal numerical postprocessing) which, to the best of our knowledge, has not yet been achieved.

In the next section, we outline the theoretical foundation of quantum tomography with PNR measurements. In section 3, we describe the TES detector and the acquisition and processing of PNR signals. In section 4, we describe the quantum tomography experimental setup and present the measurement results. In section 5, we discuss experimental limitations (losses), and we conclude. 

\section{Quantum tomography by  counting photons~\cite{Banaszek1996}}

The Wigner function~\cite{Wigner1932} of a single mode of the quantum electromagnetic field of density operator $\rho$ can be written~\cite{Leonhardt1997} as
\begin{align}
W(q,p) &= \frac{1}{2\pi}\int^{\infty}_{-\infty} e^{iyp}\bra{q-\frac{y}{2}} {\rho} \ket{q+\frac{y}{2}}dy,
\end{align}
where $\ket{q\pm y/2}$ belong to the amplitude quadrature eigenbasis. If we write ${\rho}=\sum_{n,n'}{\rho_{nn'}}\ket{n}\bra{n'}$ in the Fock basis and use the Hermite polynomial expression of the (here, amplitude) quadrature-Fock Clebsch-Gordan coefficients~\cite{BarnettRadmore1997}, 
\begin{equation}
\langle q | n\rangle = \pi^{-\frac14}(2^{n}n!)^{-\frac12}\, e^{-\frac{q^{2}}2}\,H_{n}(q),
\end{equation}
where $H_{n(q)}$ is the Hermite polynomial of order $n$. It is then straightforward to obtain the following remarkable relation from the orthogonality of Hermite polynomials
\begin{align}
W(0,0) &= \frac{1}{\pi} \sum_n (-1)^n \rho_{nn}
\end{align}
(where the right-hand side can be construed as the expectation value of the photon-number parity operator $\Pi = \sum_n (-1)^n \ket n\bra n$).
Since $\rho_{nn}$ is the probability to count $n$ photons in $\rho$, it is therefore clear that the value of the Wigner function at the origin can be obtained directly from the statistics of ideal PNR measurements. Banaszek and W\'odkiewicz then proposed to access the Wigner function in the rest of the quantum phase space by simply displacing the quantum state $\rho$ by $\alpha=(q+ip)/\sqrt2$. In practice, this can be achieved by combining the quantum signal with the coherent state $\ket\alpha$ of a well stabilized laser beam, at a beam splitter of field transmission and reflection coefficients $t$ and $r$, respectively ($r^{2}+t^{2}=1$). If the beam splitter transmits the quantum mode to be measured and reflects the coherent mode (for example), then the Wigner function measured by PNR statistics at the beam splitter's output will be~\cite{Banaszek1996}
\begin{align}
W_\text{out}(0,0)= \frac1T W\left(\frac rtq,\frac rtp; -\frac rt\right),
\end{align}
where the function $W(q,p;s)$ on the right-hand side is the $s$-ordered quasiprobability distribution~\cite{Cahill1969}, which coincides with the Wigner function for order parameter $s=0$. Hence, by choosing $r\simeq 0$ and by scanning $\alpha=(q+ip)/\sqrt2$ in the $(q,p)$ phase space, we can measure the complete Wigner function of $\rho$.

Note that the single-TES restriction of the maximum number of measurable photons (here 5 photons at 1064 nm) entails a  restriction of the  Hilbert space to the Fock-state basis $\{\ket0,\ket1,\ket2,\ket3,\ket4,\ket5\}$. As noted in Ref.~\citenum{Banaszek1999a}, when the average detected photon number approaches the cutoff limit of the detector, statistical errors increase drastically. Therefore this quantum tomography method requires states with negligible probabilities of measuring photon numbers higher than 5. However, this is not a sharp limitation: with 8 independent TES channels, our system could achieve, in principle, PNR detection up to 40 photons, and sophisticated data processing methods, mentioned in the next section, allow to push that limit even farther away.

\section{Photon-number-resolving detection setup}

\begin{figure}[t]
	\includegraphics[width=.7\columnwidth]{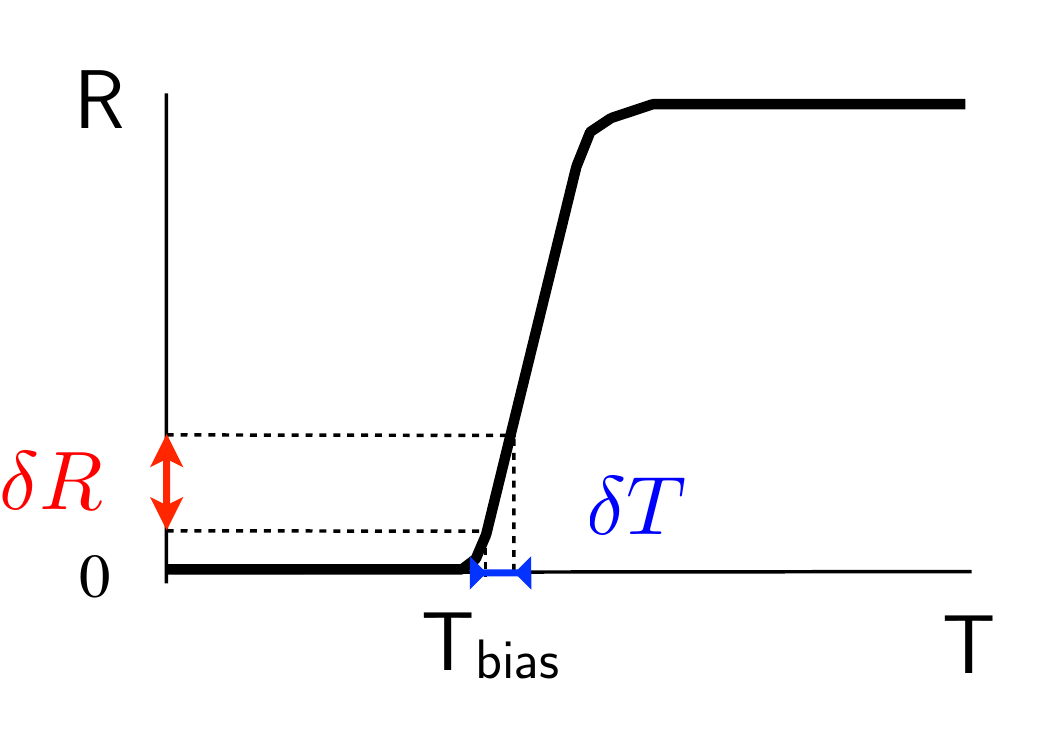}
	\vglue -.2in
	\caption{Principle of operation of the TES. When the TES is temperature-biased at the edge of the superconducting transition $T_\text{bias}$ --- or on the steep transition slope --- any temperature variation $\delta T$ due to photon absorption is translated into a measurable resistance change $\delta R$.}
	\vglue -.2in
	\label{fig:TES}
\end{figure}

Our TES system contains 8 fiber-coupled thin-film tungsten devices fabricated at NIST, optimized for detection at 1064 nm~\cite{Lita2008}. 
The TES devices were cooled by a cryogen-free adiabatic demagnetization refrigerator and temperature stabilized at 100 mK. The TES detector is voltage-biased~\cite{Irwin1995}, and self-heats into the superconducting transition illustrated in \fig{TES}.  When a photon is absorbed, the energy of the photon is thermalized in the electrons of the TES and there is a small temperature rise that causes a small measurable increase in the resistance of the TES.  The change in resistance causes a change in current flowing through the device which is measured using a SQUID amplifier system.  

Two-photon absorption causes a larger temperature change and therefore a larger signal than single-photon absorption, and this results in photon-number resolution. 
 
{Since each quantum of light that is absorbed within the cooling time raises the temperature, and therefore the resistivity, of the device, and since that resistivity change constitutes the exploitable TES signal, } the maximum measurable photon number is ultimately determined by the amplitude of the steep transition slope (\fig{TES}): upon reaching its top, the TES will saturate if additional photons are absorbed. There exist, however, methods to cope with such undesirable conditions as TES saturation: on the one hand, the cooling time would still provide information about the photon number in such an optical pulse, if no more photons were impinging until cooling was complete~\cite{Figueroa-Feliciano2000}. Moreover, the saturation does not completely erase all photon-number information (unlike the schematic plot of \fig{TES}, the resistance does retain a weaker dependence on temperature in the normal conducting regime) and a recent, more sophisticated analysis~\cite{Levine2012} can also yield higher photon number statistics into the saturation regime.

\begin{figure}[t]
	\includegraphics[width=\columnwidth]{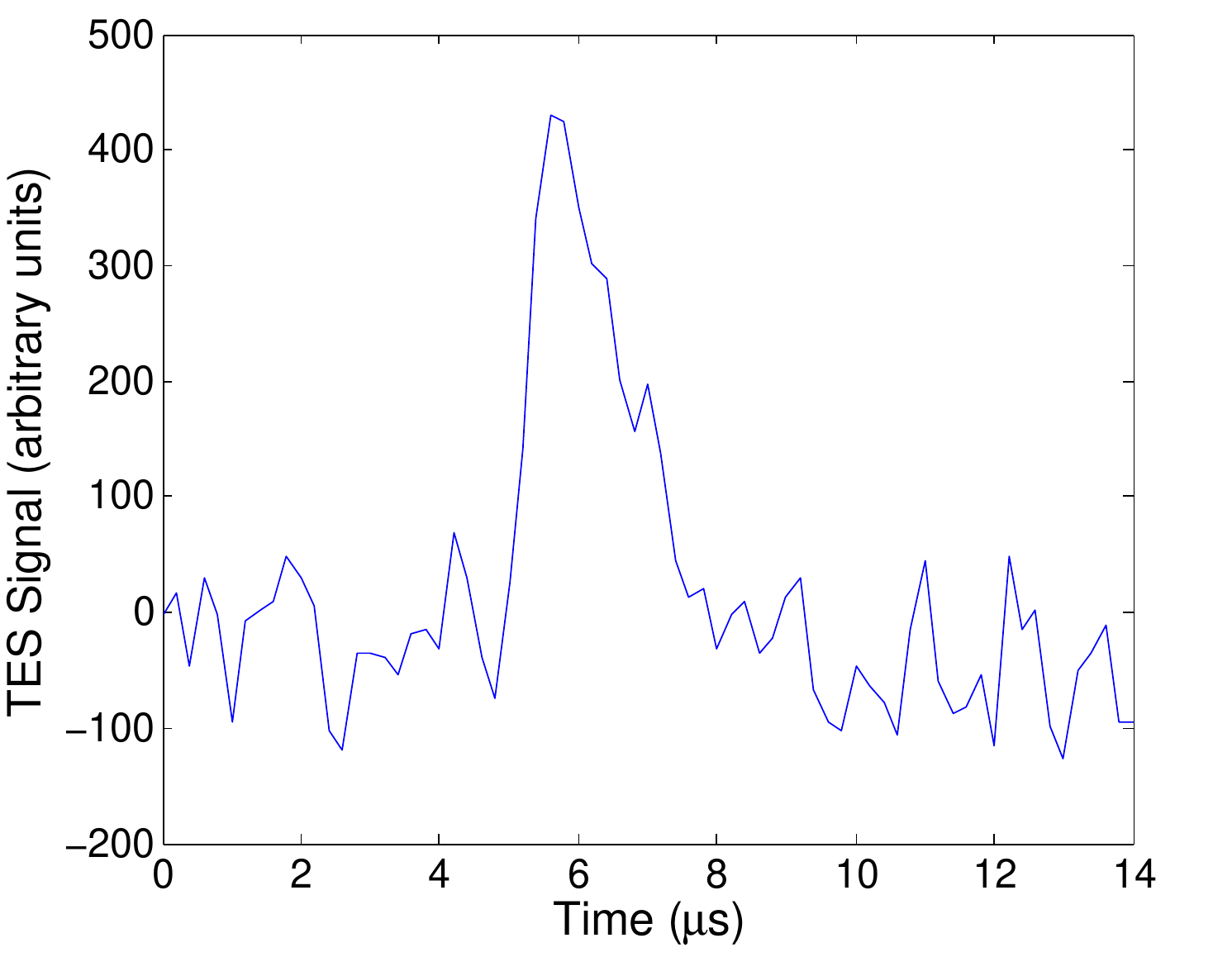}
	\caption{An example of raw TES data showing a single photon detection event. The detection peak can be clearly distinguished from the noise.}
	\label{fig:peak}
\end{figure} 
\begin{figure}[t]
	\includegraphics[width=\columnwidth]{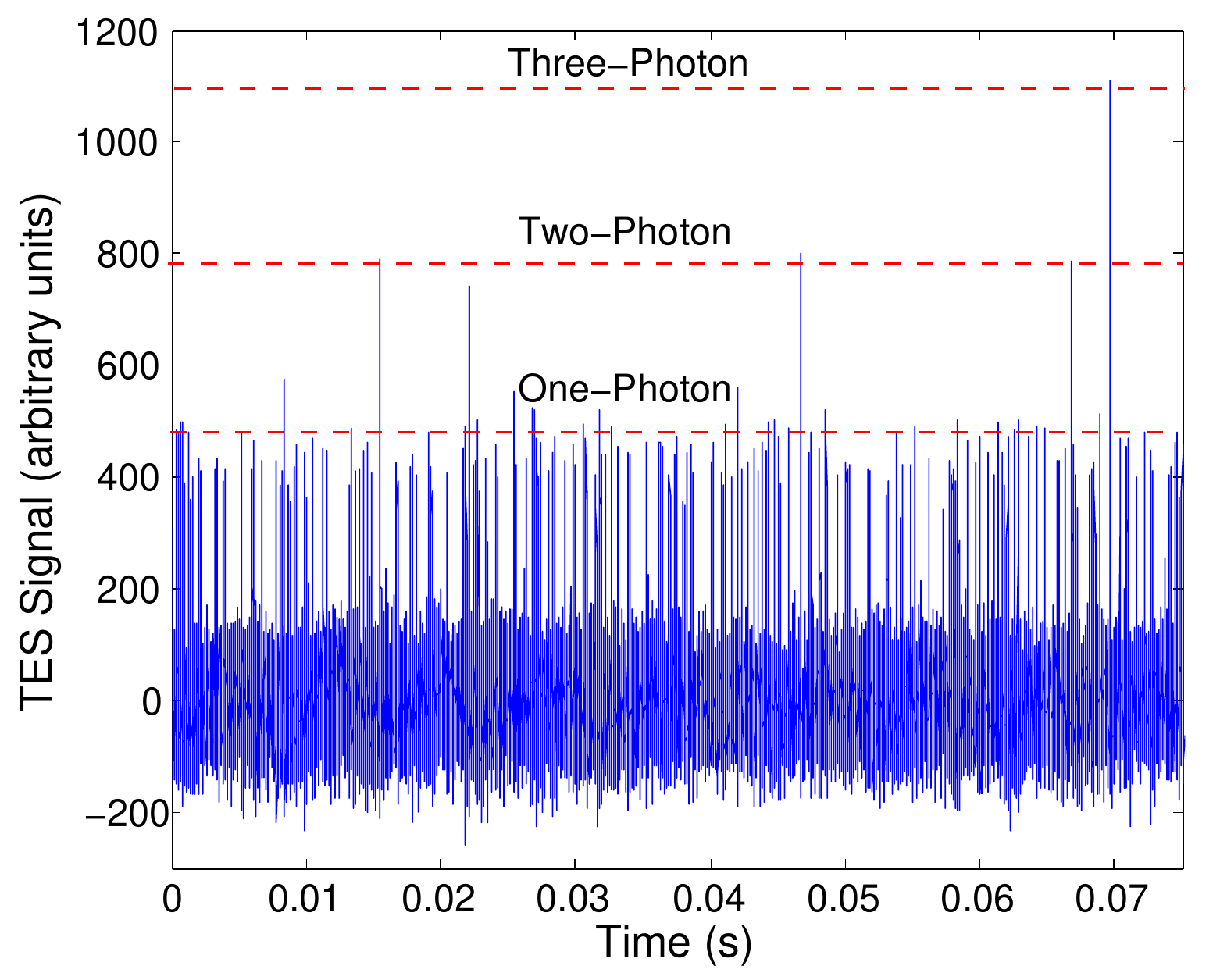}
	\caption{Raw TES data for a CW beam showing distinct one-, two-, and three-photon peaks. The corresponding signal levels are indicated by the red dashed lines.}
	\label{fig:TESsignal}
\end{figure} 

\subsection{Signal acquisition}

The TES signal comprised a rising edge of about 700 ns, corresponding to the response time of the TES detection chain, followed by a cooling decay tail of a couple $\mu$s. \Fig{peak} displays a typical example.
Note that the detector is still active during the cooling tail and that there is no dead time as long as the TES is on the transition slope. 

The TES signal was then sampled at a sampling rate of 5 Ms/s. The data was saved in packets of $2^{22}$ points.  Each point was saved on the computer as a 16-bit integer, but only 14-bits were useful from the digitizer.  Therefore, the size of each packet was 8 MBytes. This process could be repeated if necessary to join multiple data packets. Each data packet corresponded to 0.84 s of uninterrupted data. In this experiment, all Wigner-function measurement data was exactly 1-packet long and contained no dead time.

\begin{figure*}[t]
	\includegraphics[width=1.5\columnwidth]{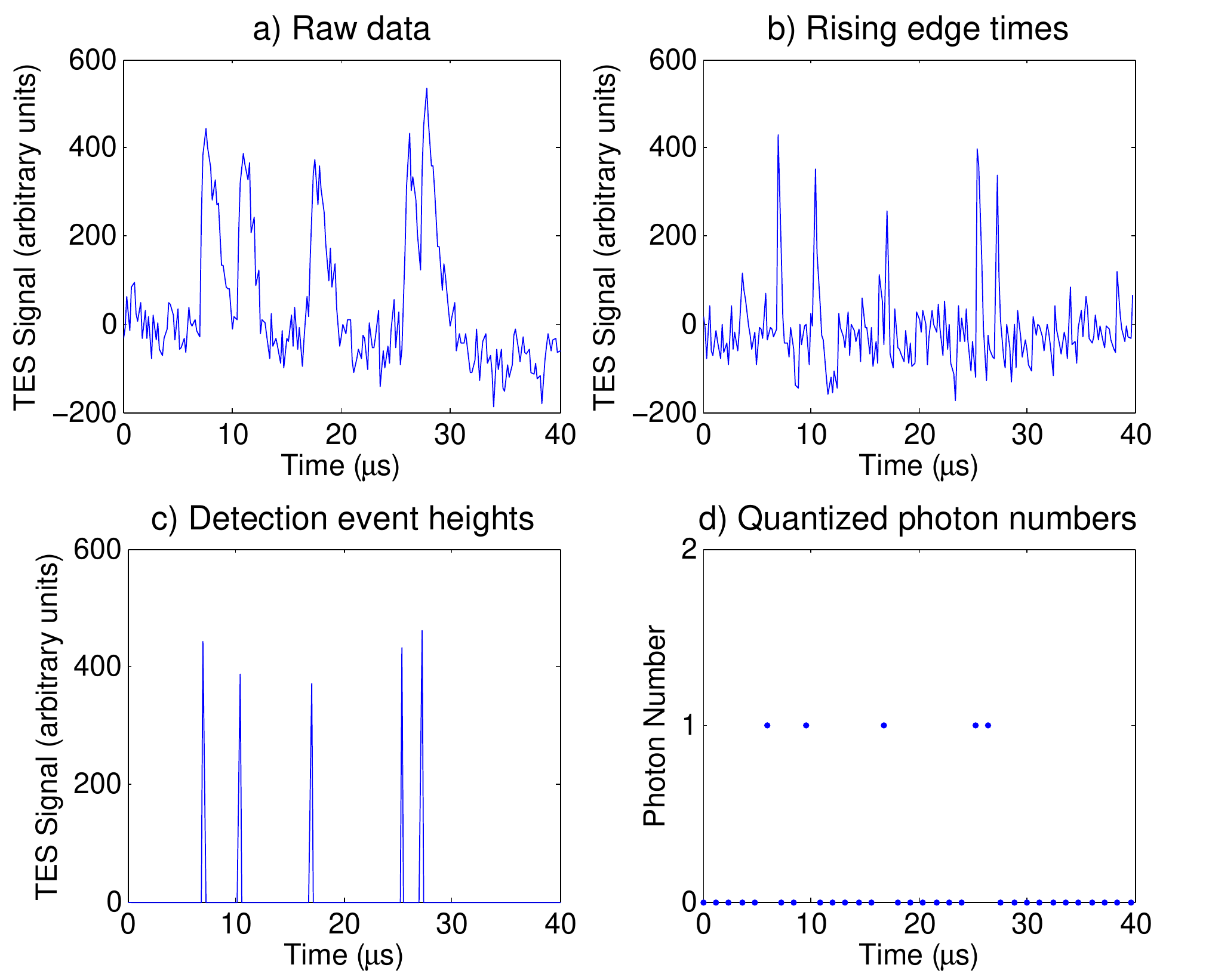}
	\caption{Steps in the processing of raw analog data into quantized photon-count data. (a), a small sample of the raw data shows 3 single peaks and 1 ``piled up'' double peak, where the last photon was detected before the detector was cooled back the nominal bias temperature. (b), rising edge detection results. The procedure correctly detected 5 rising edges. (c), encoded signals, made of the photon detection times along with the value of the maximum peak height within 1.2 $\mu$s of each event. (d), final quantized photon-counts determined using the thresholds defined in the histogram of peak heights in \fig{histogram}.}
	\label{fig:pileups}
\end{figure*} 

\subsection{Signal processing and photon ``pileups''}

In all experiments, continuous-wave (CW) optical fields were measured and all TES signals were derived from continuous photon streams. \Fig{TESsignal} displays an example of PNR detection with the TES, over a longer time range than \fig{peak}.
In the data of \fig{TESsignal}, the photon flux was kept low enough that all detection peaks were separated by more than the TES cooling time. In other words, there was no ``photon pileup'' event in which another photon impinged on the TES shortly after a first photon, while the signal is still on the decaying tail. An example of pileup is the rightmost (double) peak in \fig{pileups}(a).
However, one can still observe small fluctuations in the detection peak heights in \fig{TESsignal}, due to noise in the readout electronics. In order to achieve accurate photon counting, including in the presence of pileups, we adopted the following procedure, whose steps are illustrated in \fig{pileups}.

\begin{figure}[t]
	\includegraphics[width=\columnwidth]{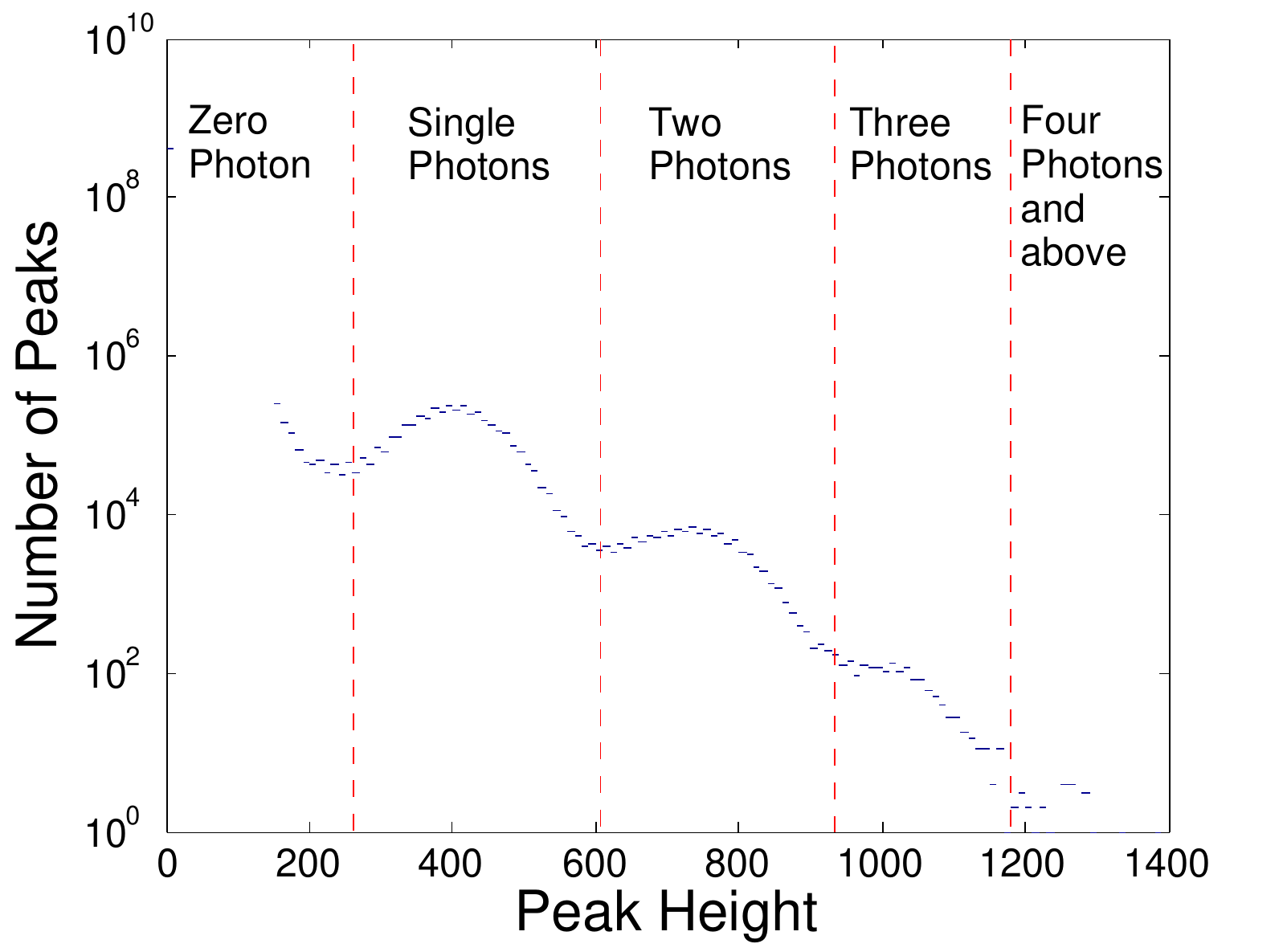}
	\caption{Histogram of peak heights from the sample of \fig{pileups}(a). The histogram bins define the photon-number quantization thresholds.}
	\label{fig:histogram}
\end{figure} 

\begin{figure*}[t]
\includegraphics[width=1.5\columnwidth]{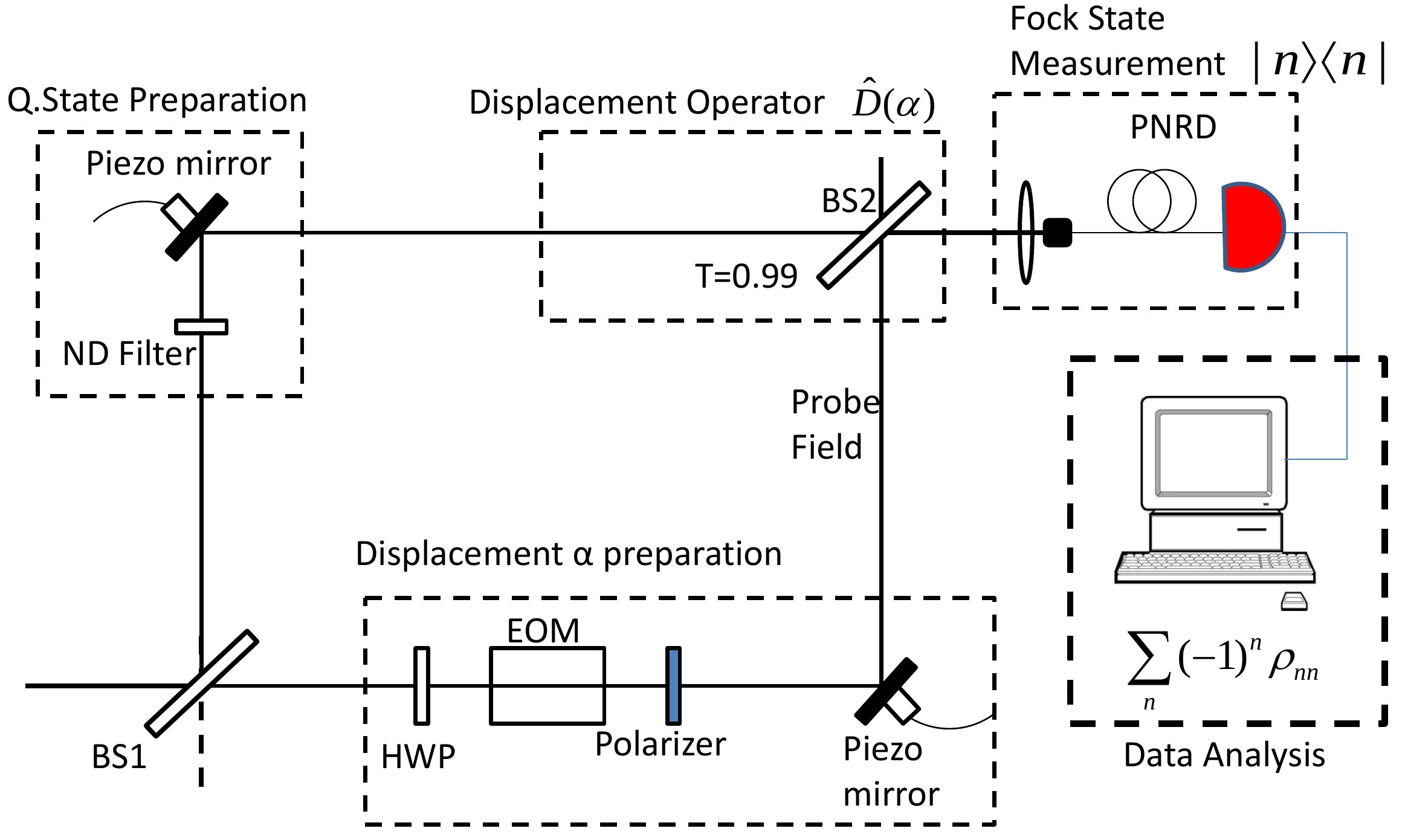}
\caption{Experimental setup for the quantum tomography of a coherent state using a coherent probe field and photon number measurements. The leftmost and upper optical path is that of the quantum state to be measured. The lower and rightmost optical path is that of the ``local oscillator'' whose field generates the phase-space displacement. BS: beam splitter; EOM: electro-optic modulator; HWP: halfwaveplate; ND: neutral density.}
\label{fig:setup}
\end{figure*}

First, we identified each detection event by finding rising edges in the signal. A rising edge is defined as a detection event if it rises at least $40\%$ of the average height of the single photon above the mean noise level. This threshold is set manually during the calibration process. For example, in \fig{peak}, the threshold would typically be at 200 arb.\ units. The starting time of each detection event is recorded. \Fig{pileups}(b) displays the event times corresponding to the signal in \fig{pileups}(a). The algorithm then stores the maximum signal in the $1.2\ \mu$s following each starting time and this maximum is stored as well, see \fig{pileups}(c). The TES response time is thus $1.2\ \mu$s. Hence, if two photons were absorbed with $1.2\ \mu$s of each other, they would be counted as one two-photon event, not as two one-photon events; this is determined in the final, quantization stage. First, we form the histogram of recorded signal heights, displayed in \fig{histogram},
where we can clearly see three well-separated peaks (the ``zero-photon'' area is likely due to blackbody radiation). From this histogram, we can now define the photon-number quantization thresholds. Using these thresholds, the quantized photon-number signals can be obtained and are displayed in \fig{pileups}(d). Note that the rightmost pileup peak is resolved here and accounted for as two one-photon events.

\subsection{Detection efficiency}

The detector efficiency is an important factor in accurate state reconstruction. Losses not only change the state being measured but also introduce noise in the statistics. Detailed analysis of the effect of loss can be found in Refs.~\citenum{Banaszek1996,Banaszek1999a} and references therein. Highly nonclassical states are very strongly affected and quickly lose their characteristics. {For example, squeezed states become less squeezed under the action of losses.} Coherent states, however,  {remain coherent states   and only see their amplitude decrease}. A detailed treatment \cite{Banaszek1999a} shows that losses only decrease the observed peak of the coherent state Wigner function, but preserve its Gaussian nature {and width}. 

In this work, we restricted ourselves to coherent states. Our preliminary characterization of the overall system detection efficiency of our TES to be at least 70\% and as high as 90\%.

\begin{figure*}[t]
     \includegraphics[width=0.62\columnwidth]{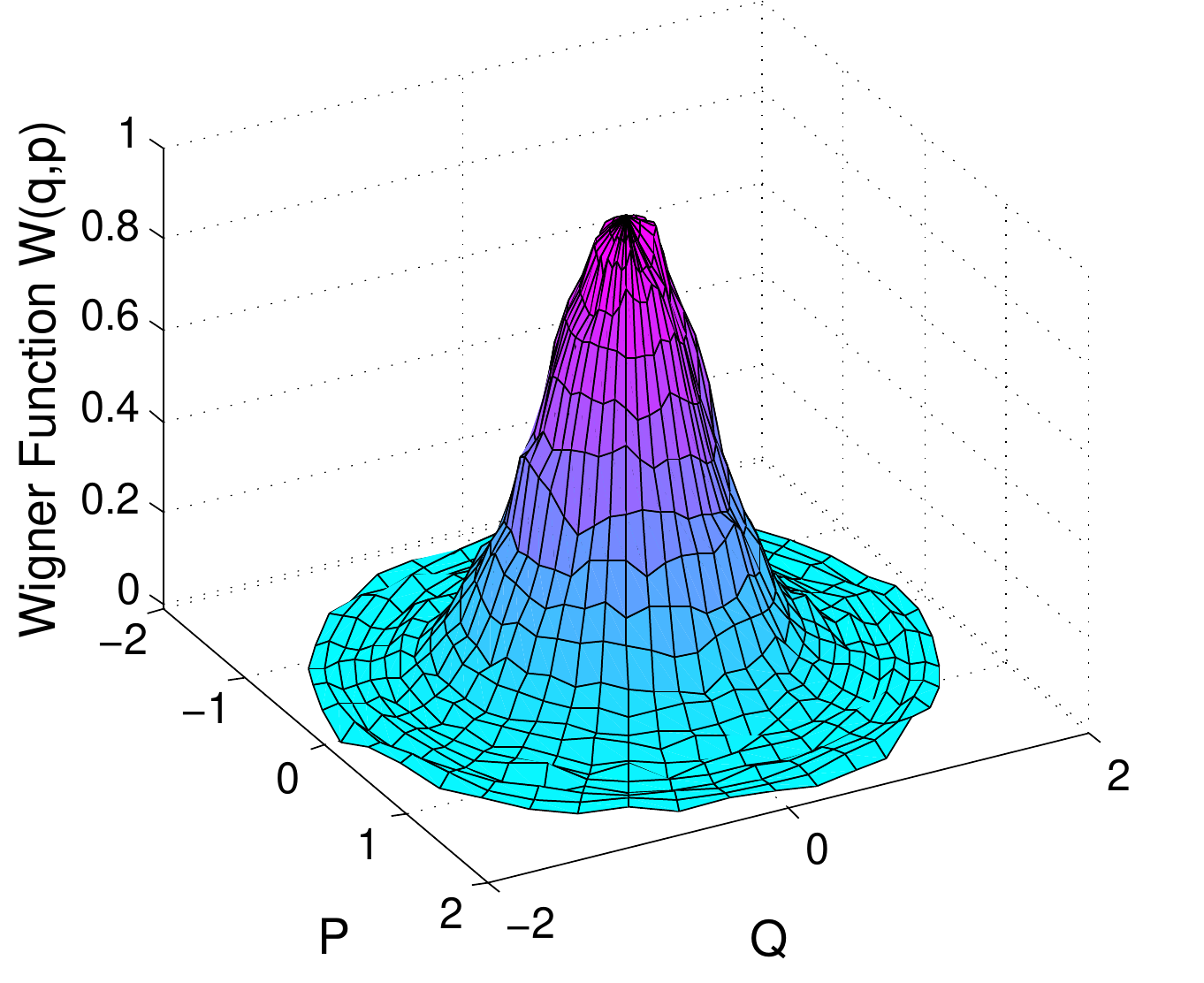}
     \includegraphics[width=0.62\columnwidth]{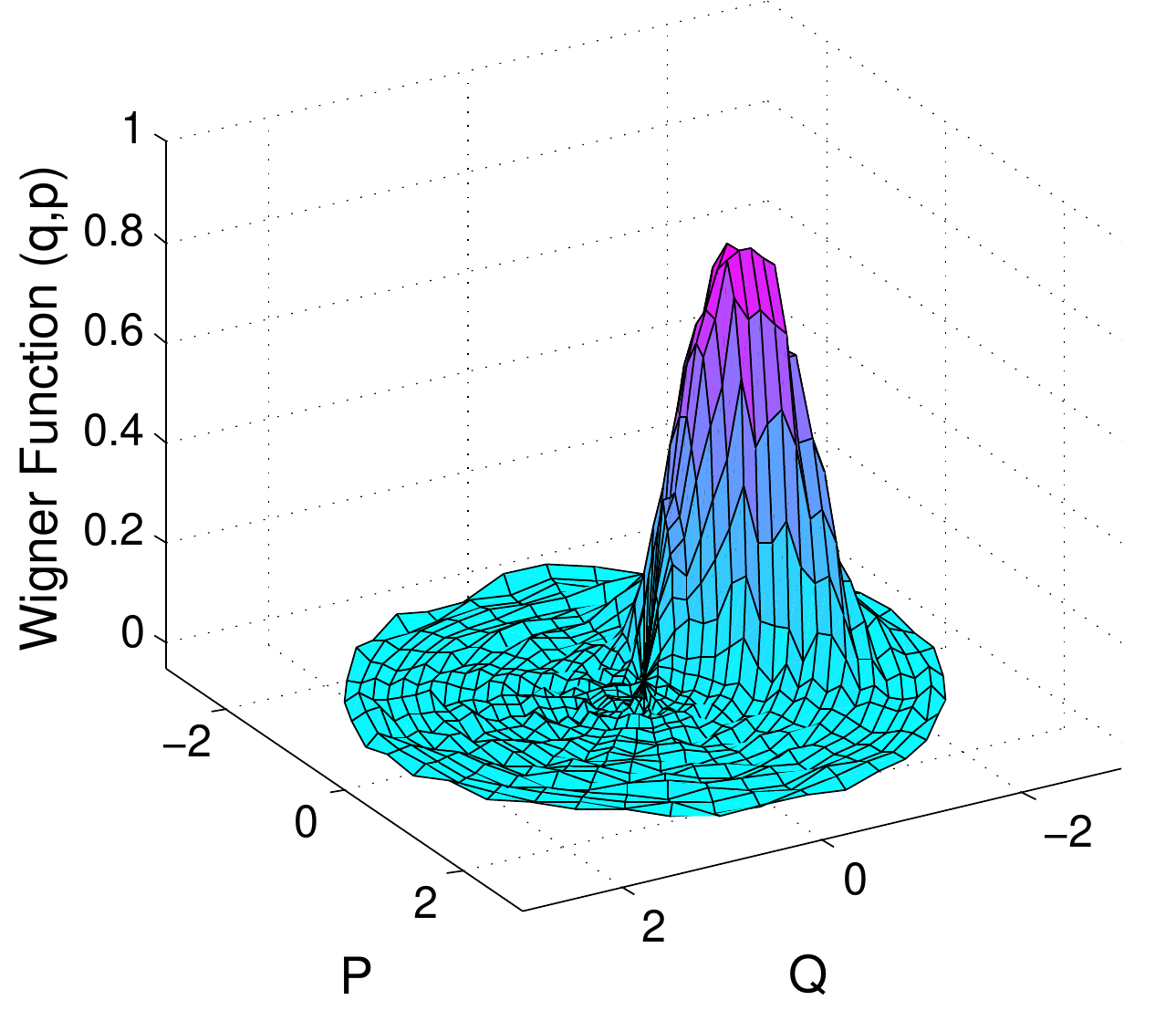}
     \includegraphics[width=0.62\columnwidth]{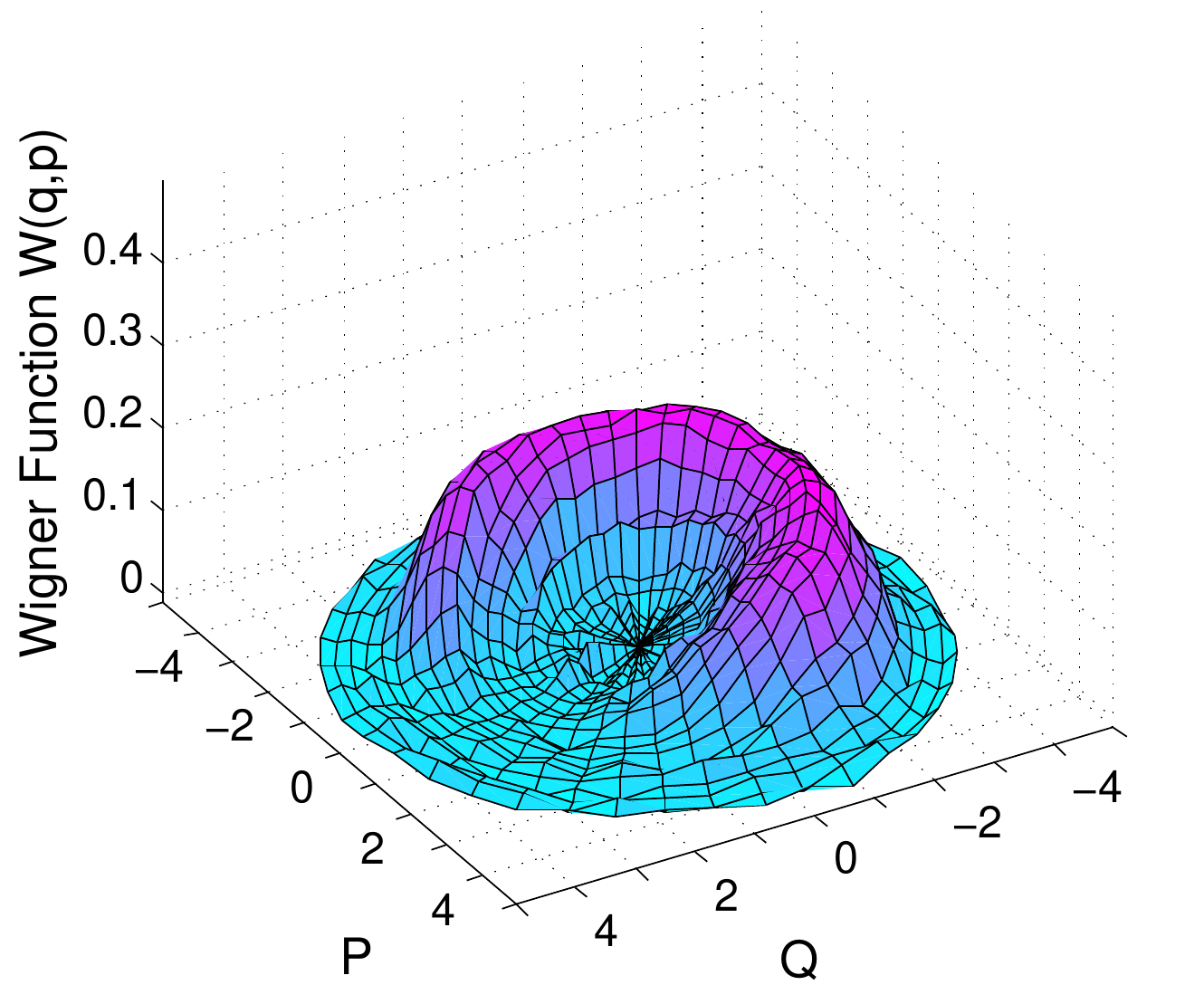} \\
     \includegraphics[width=0.62\columnwidth]{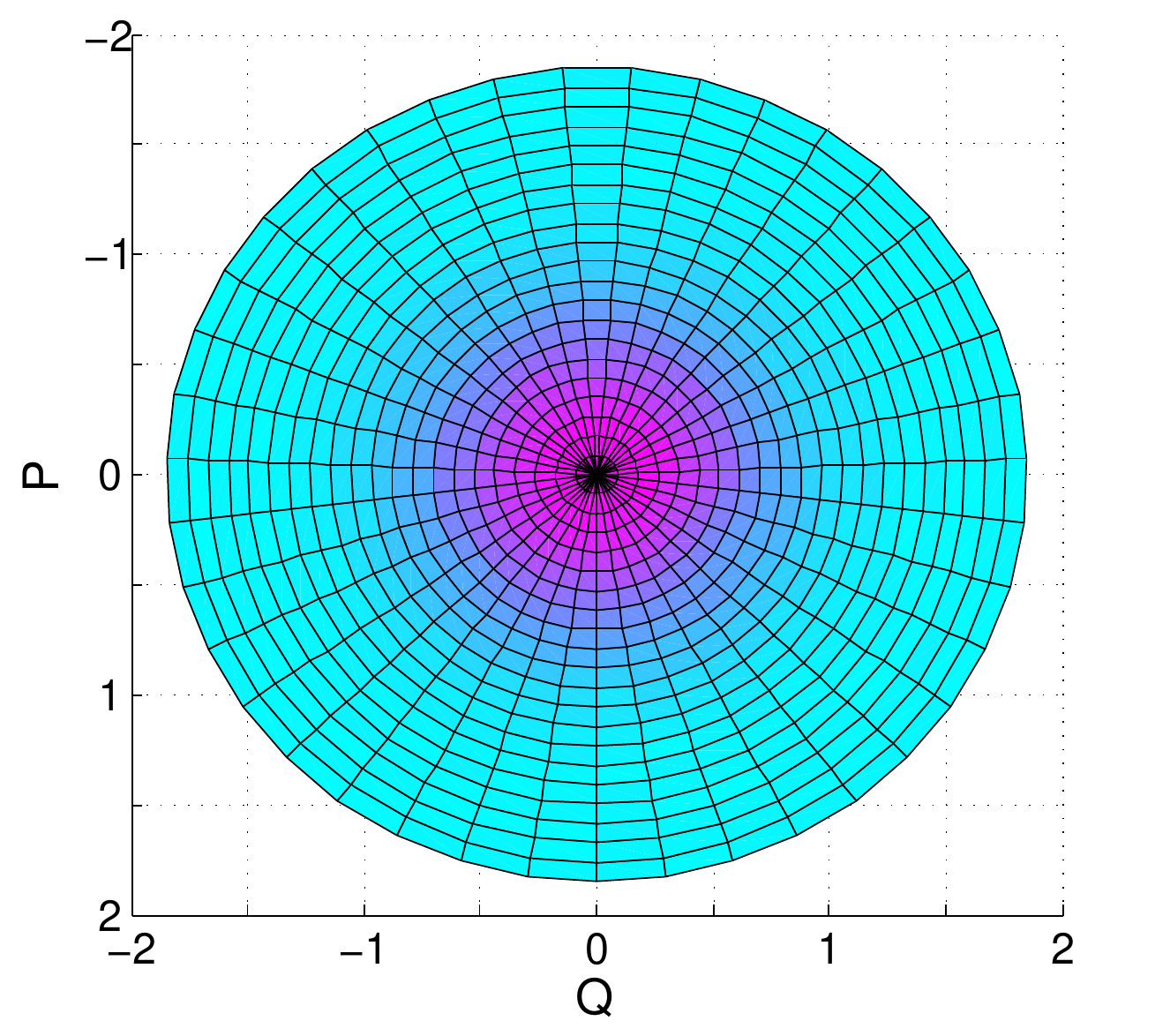}
     \includegraphics[width=0.62\columnwidth]{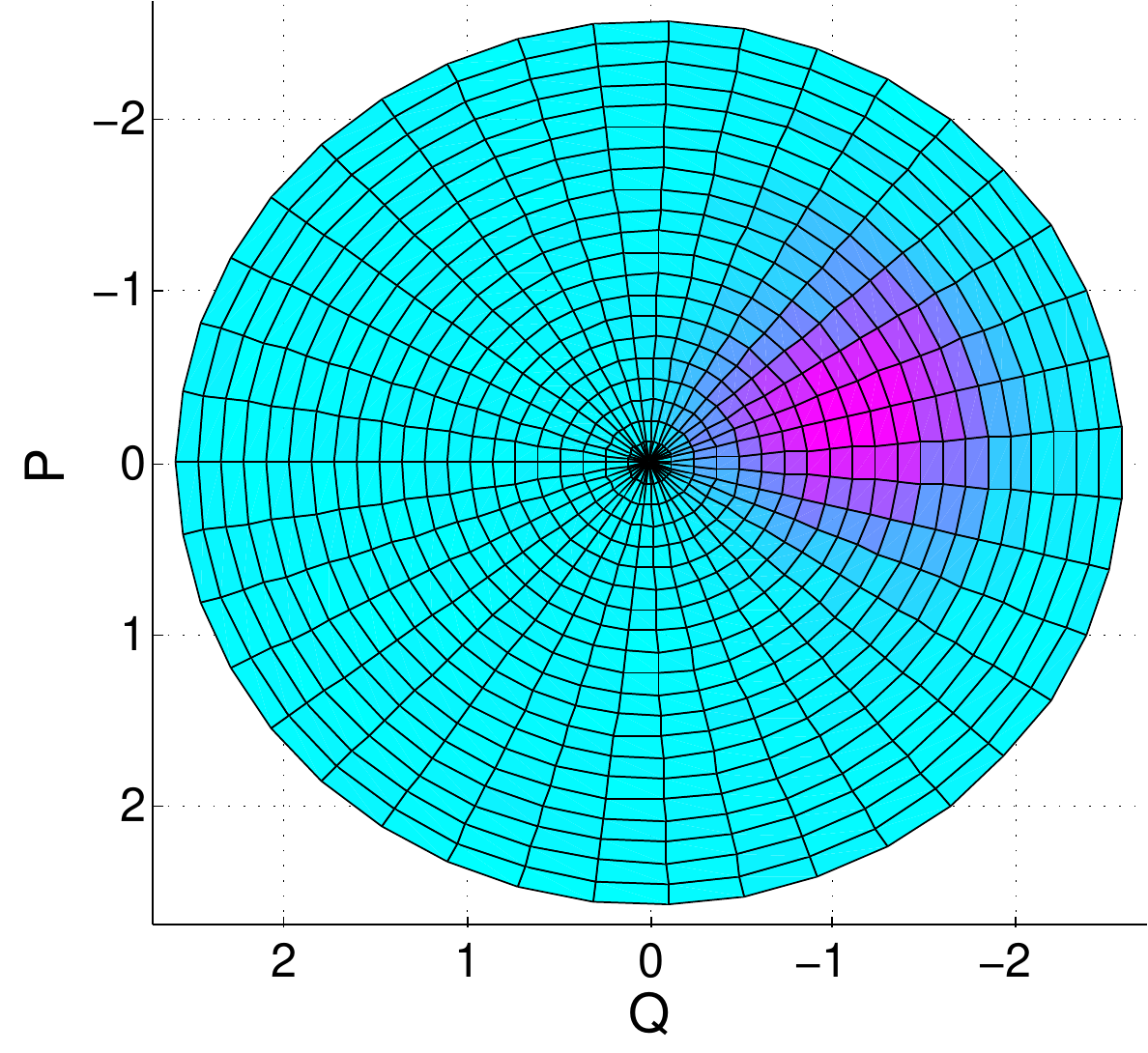}
     \includegraphics[width=0.62\columnwidth]{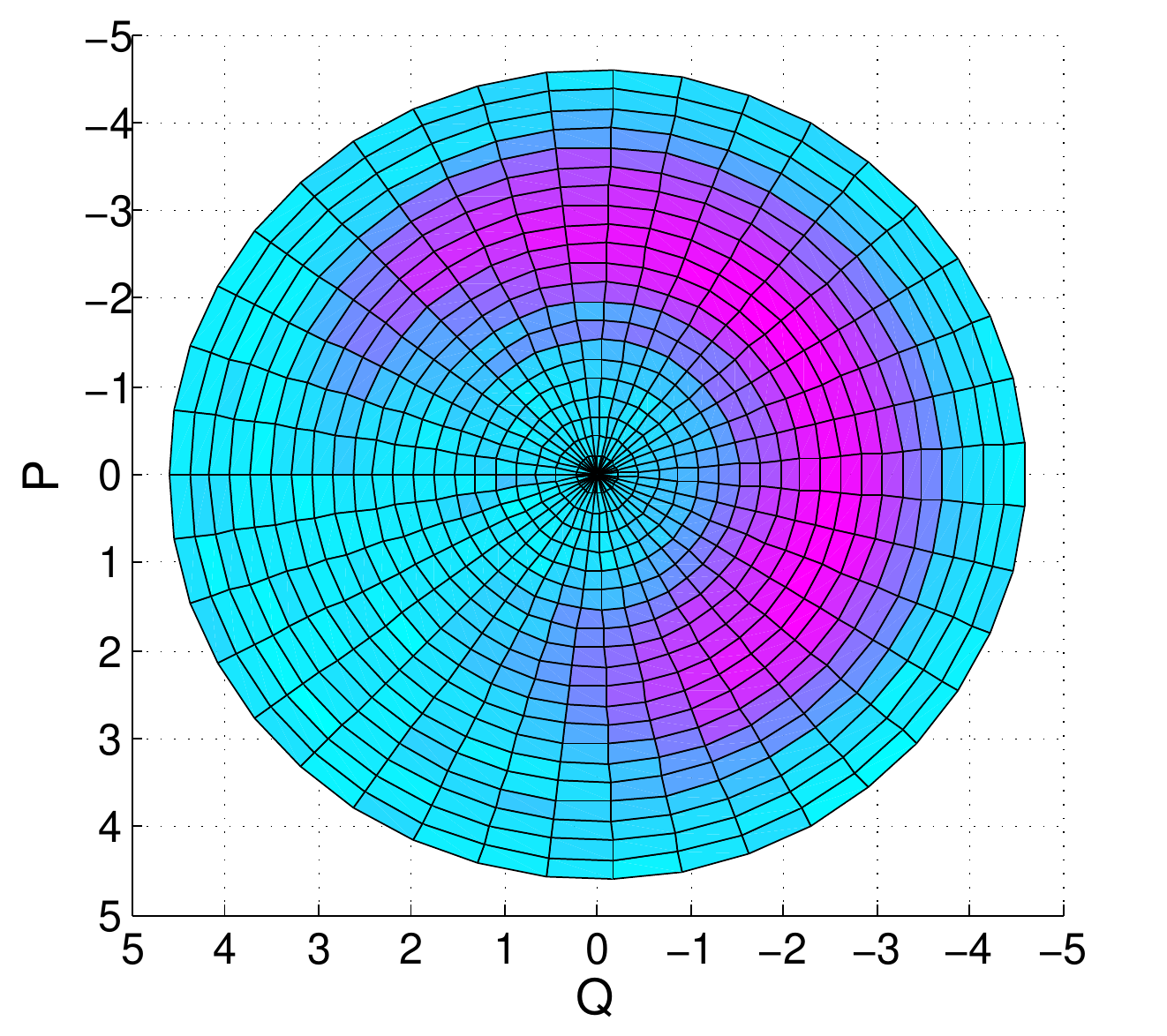}
\caption{The measured Wigner functions and contours of, left, vacuum; center, a coherent state; right, a phase-diffused coherent state. All plots are interpolated for 22 amplitude points and 40 phase points.} 
\label{fig:Wignerplots}
\end{figure*}

\section{Quantum tomography setup and procedure}

\subsection{Setup}

The experimental setup schematic is depicted in \fig{setup}. The whole experiment was set up on a 24 inch-thick floating optical table and all optical paths were protected from air drafts by acrylic plastic enclosures. All light was emitted at 1064 nm by a monolithic Nd:YAG laser, of high intrinsic stability (1 kHz FWHM linewidth). The optical mode was coupled to single-mode fibers for 1550 nm light, antireflection (AR) coated at 1064 nm, by way of aspheric lenses, also AR-coated, and a 5-axis fiber aligner. These fibers entered the cryostat via throughputs and were then directly coupled to the superconducting detectors by silicon micromachined self-alignment.~\cite{Miller2011}. As mentioned above, measuring the Wigner function requires, besides PNR detection, quantum state displacement over the whole region of interest of the phase space. The displacement operator was implemented by interfering the signal field with a local oscillator (LO) coherent field at a nearly fully transmitting beam splitter. {The justification of this arrangement is easy to conceive when first considering the operator for a beam splitter of field reflection coefficient $r$, onto which 2 fields $a$ and $b$ are impinging,
\begin{equation}
e^{2i\arccos r (\d ab+a\d b)}.
\end{equation}
If field $a$ is in a coherent state $\ket\alpha$, the operator may be written
\begin{equation}
e^{2i\arccos r (\alpha^{*}b+\alpha\d b)},
\end{equation}
which coincides with a displacement operator. The final step is to choose $r\simeq 1$ or $r\ll1$ in order to ensure that the quantum signal $b$ is not actually split in a significant way.}
The interference visibility of the signal and LO fields was $v=98\%$. The amplitude $|\alpha|$ and phase $\arg(\alpha)$ of the transmitted coherent field were respectively varied using an amplitude electro-optic modulator (EOM) and a piezotransducer-actuated mirror. The EOM we used was a home-made device built out of an $X$-cut, 20 mm-long $\rm RbTiOAsO_{4}$ (RTA) crystal, which was temperature controlled to about 1 mK by a commercial temperature controller. The voltages applied to EOM and piezo mirrors were generated by low-noise, high-voltage drivers controlled by the analog output ports of a lock-in amplifier. The lock-in amplifier was computer-controlled to output desired voltages through its auxiliary A/D outputs, which have 1 mV resolution, $\pm$10 V range, and under 100 $\mu$s settling time. However, the fastest switching time we observed was 10 ms, which is likely a remaining limitation of the interface rather than the limit of the lock-in amplifier itself.

\begin{figure*}[t]
\begin{center}
\includegraphics[width=2\columnwidth]{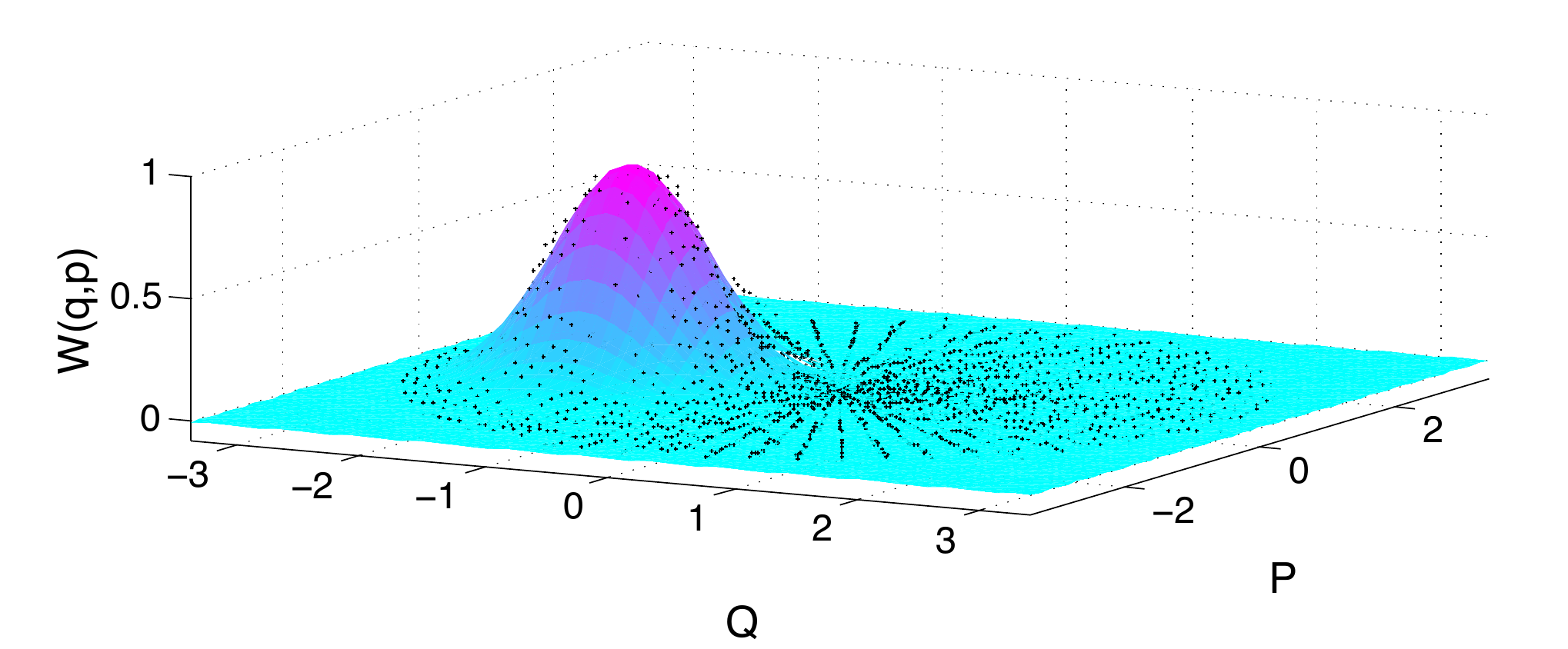}\\
\includegraphics[width=2\columnwidth]{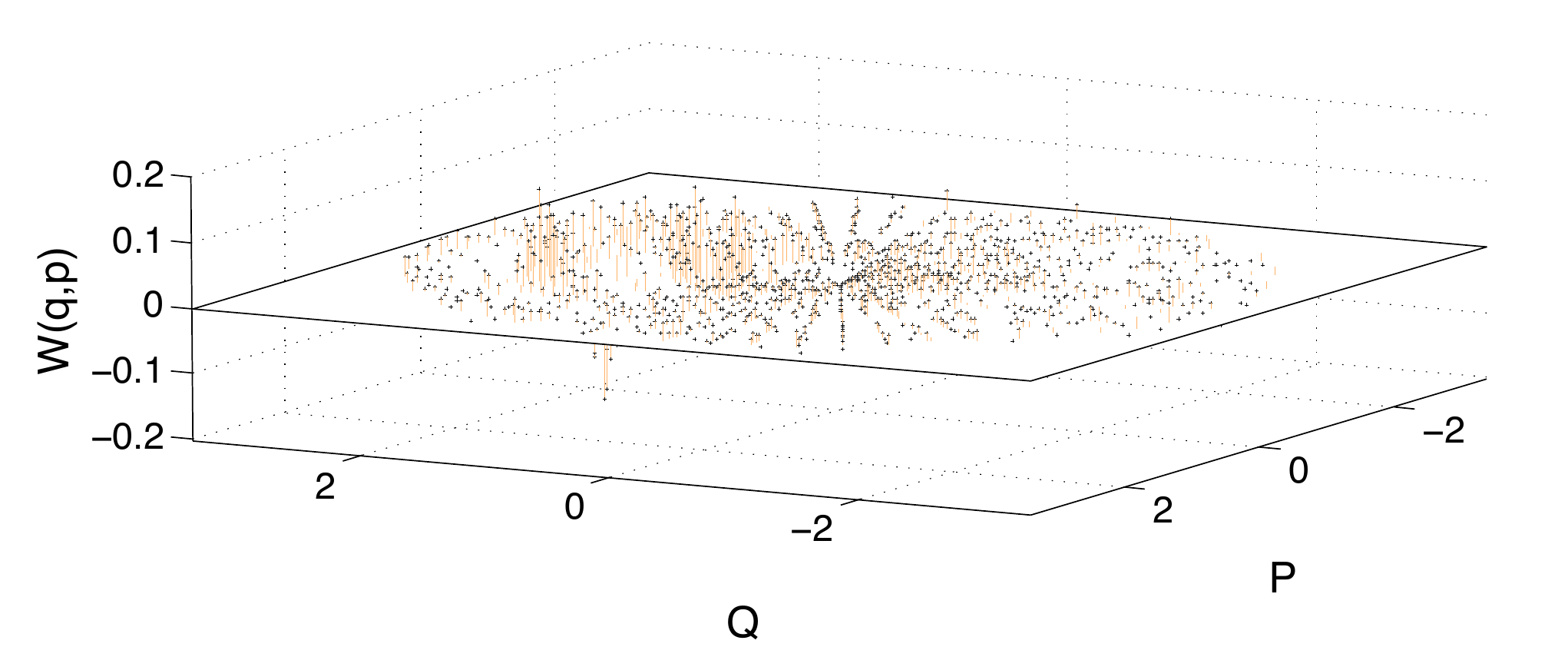}
\caption{Gaussian fit (top) and residuals (bottom) of the coherent-state tomographic data. The fit function was $W_\text{fit}(q,p) = a+ b\, \exp[-m|(q-q_0)^2 +(p-p_0)^2|]$. The correlation coefficient was $R^2=0.966$.}
\label{fig:fit}
\end{center}
\end{figure*} 

\subsection{Procedure}

The amplitude $|\alpha|$ of the displacement field was varied in 40 steps while its phase $\arg(\alpha)$ went through 60 steps from 0 to $2\pi$. At each point ($|\alpha|$,$\arg(\alpha)$) of the polar scan the TES signals were processed to yield photon-counting statistics, hence photon-count probabilities and also parity measurements. Phase space scans consisted of sampling a sequence of circles with increasing radius, since changing the voltage applied to the EOM tuning $|\alpha|$ required a settling time of the order of 2 s, whereas $\arg(\alpha)$ could be scanned much faster, as the phase modulator driver had much higher, 10 kHz bandwidth. For each point in phase space, the TES output was digitized and processed to obtain the photon statistics in real-time. The parity measurement was subsequently calculated from the statistics and saved.

\section{Results and analysis}

\subsection{Reconstruction}

In \fig{Wignerplots}, we plot the measured Wigner functions of the vacuum, of a weak coherent state, and of a phase-diffused statistical mixture of coherent states. The phase diffusion was  obtained by applying a 100 Hz sine waveform to the piezo mirror. The radial coordinate was obtained from the average number of photons detected for the blocked signal path. Thus the graphs were parameterized with the complex variable $\beta = \sqrt{\eta}\alpha$, where $\alpha$ is the probe field reflected at the beamsplitter and $\eta$ is the detection efficiency. 

Each 0.84 s data packet (see above) was divided into bins of size $\tau=0.1$ ms long, which amounted to about 8400 bins per point in phase space. The bin duration defined the temporal envelope of the measured mode. Like in Banaszek et al.'s original experiment, this was shorter than the laser's coherence time (here, 1 ms) but of no consequence in the case of coherent states.

 \subsection{Verification}
We investigated the weak coherent state case to verify the accuracy of our state reconstruction. The theoretical Wigner function of a coherent state $\ket{\alpha_0}$ is the well known
\begin{align}
W(\alpha) 
&= e^{-2|\alpha-\alpha_0|^2 }.
\end{align}
However, a more realistic analysis~\cite{Banaszek1996,Banaszek1999a} takes into account sub-unity detector efficiency, losses on the quantum signal and the nonideal visibility of the signal-LO interference at the displacement beamsplitter BS2. The measured Wigner function is given by
\begin{align}
W(\beta) = e^{-2\left|\beta-\sqrt{V\eta} t\alpha_0\right|^2-2(1-V)\eta t^{2}|\alpha_0|^2},
\end{align} 
where 
\begin{equation}
V=\frac{v}{2-v}
\end{equation}
is a measure of the overlap of signal and LO.

In this experiment, we had $v=0.98$, $\eta=0.72$, $t^{2}=0.99$, yielding $V = 0.97$.
We recall that the measured probe field in \fig{Wignerplots} is already $\beta$ and the measured signal is $|\beta_0|^2=\eta|\alpha_0|^2$=2.553, therefore $|\beta_0|$=1.597.
Thus, our theoretical Wigner function was
\begin{align}
W_\text{fit}(\beta) = 0.867\,e^{-2|\beta-1.597|^2}, \label{eq:W}
\end{align}
which we compared to a fit of the observed data in \fig{Wignerplots}. The fit is plotted in \fig{fit} and the results are presented in \tb{T}.
\begin{table}[t]
\caption{Fit results for model function \\ $W_\text{fit}(q,p) = a+ b\, \exp[-m|(q-q_0)^2 +(p-p_0)^2|]$. {Note that only the modulus $|\beta_{0}|$ was fitted, as its argument was arbitrary --- if static --- in the experiment.} }
\[
\begin{array}{|c|c|c|}
 \hline
\text{Coefficients} & \multicolumn{1}{c|}{\text{Fit}} 
& \multicolumn{1}{c|}{\text{Theory (\ref{eq:W})}} \\
 \hline
a &  \bf 0.000(2)  & \bf 0\\
 b &   \bf   0.877(10)  &\bf 0.867\\
 m &   \bf    1.72(3)  & \bf 2\\
 p_{0} &      0.248(6)  & \\
 q_{0} &     1.532(6)  & \\
 \sqrt{q_{0}^{2}+p_{0}^{2}} &  \bf   1.552(6)  & \bf 1.597\\
 \hline
\end{array}
\]
\label{tab:T}
\end{table}
Note that the actual phase angle of the coherent state [$\arctan(p_{0}/q_{0})$] is not relevant here, even though we did fit it, only the amplitude $\beta_{0}=\sqrt{q_{0}^{2}+p_{0}^{2}}$ is. Although the correlation coefficient of the fit was high ($R^2=0.966$), an inspection of \tb T and of the residuals in \fig{fit} show that the agreement is only qualitative. We believe the reason for the discrepancy lies in systematic errors in the data acquisition, which lasted for 4000 s for the complete data. While our experimental setup is intrinsically very stable, there remain likely, if small, phase drifts over such a duration, since the optical phases are not locked. We plan to reduce the acquisition time and implement higher stability of the optical paths in the future.

\section{Conclusion}

In conclusion, we experimentally demonstrated quantum tomography with PNR measurements of more than one photon. We limited this initial investigation to the loss immune coherent state and coherent-state mixture with phase noise and got reasonable agreement with expected values on key parameters. Since the detector's nonideal efficiency cannot be compensated in this method of measuring the Wigner function \cite{Banaszek1996}, it is only possible due to the recent development of high efficiency PNR detectors. In future work, we will improve the performance of this method so as to make it applicable to accurate state reconstruction of highly nonclassical states, such as squeezed and Fock states, which first requires us to further increase the system detection efficiency. As suggested by Banaszek et al.~\cite{Banaszek1999a}, this method also generalizes very naturally to multimode tomographic reconstruction, which is a promising direction of research.

This work was supported by the U.S. National Science Foundation, under grants No.\ PHY-0960047 and PHY-1206029, and by the University of Virginia.

\bibliography{Pfister} 
\bibliographystyle{osajnl.bst}

\end{document}